\documentstyle[11pt,aasms4,flushrt]{article}
\doublespace
\begin{document}
\title{Mid-Infrared Images of Luminous Infrared Galaxies in a Merging Sequence}

\author{Chorng-Yuan Hwang\altaffilmark{1,2}, K.Y. Lo\altaffilmark{1,3},
Yu Gao\altaffilmark{3,4}, Robert A. Gruendl\altaffilmark{3}, and 
Nanyao Lu\altaffilmark{5}}
\altaffiltext{1}{Institute of Astronomy and Astrophysics, Academia Sinica,
P.O. Box 1-87, Nankang, Taipei, Taiwan; hwang@asiaa.sinica.edu.tw, 
kyl@asiaa.sinica.edu.tw}
\altaffiltext{2}{Institute of Astronomy, National Central University,
Chung-Li, Taiwan}
\altaffiltext{3}{Department of Astronomy, University of Illinois, 
1002 W. Green Street, Urbana, IL 61801, U.S.A.;
gao@astro.uiuc.edu, gruendl@astro.uiuc.edu}
\altaffiltext{4}{current address: Department of Astromnomy, University 
of Toronto, 60 St. George Street, Toronro, ON M5S 3H8, CANADA}
\altaffiltext{5}{IPAC, MS100-22 Caltech, Pasadena, CA 91125, U.S.A;
lu@ipac.caltech.edu}

\begin{abstract}
We report mid-infrared observations of several luminous infrared galaxies
(LIGs) carried out with the Infrared Space Observatory ({\it ISO\/}).
Our sample was chosen to represent 
different phases of a merger sequence of galaxy-galaxy interaction
with special emphasis on early/intermediate stages of merging.
The mid-infrared emission of these LIGs shows 
extended structures for the early and intermediate mergers,
indicating that most of the mid-infrared luminosities are not
from a central active galactic nucleus (AGN).
Both the infrared hardness (indicated by the
IRAS 12, 25, and 60 $\micron$ flux density ratios) and the peak-to-total 
flux density ratios of these LIGs 
increase as projected separation of these interacting galaxies 
become smaller, consistent with increasing 
star formation activities that are concentrated to a smaller 
area as the merging process advances. These observations provide
among the first observational constraint of largely theoretically 
based scenarios.

\end{abstract}


\section{INTRODUCTION}
Luminous infrared galaxies (LIGs) emit most of their 
energy in the infrared wavelengths and are the most 
numerous extragalactic sources 
with bolometric luminosity larger than $10^{11} L_{\sun}$ in the local 
Universe (Sanders and Mirabel 1996). 
The energy source of 
the infrared emission in these LIGs 
is a fundamental issue in the studies of this subject.
Most LIGs are found to be interacting galaxy systems, suggesting  
that collisions and mergers are the major mechanisms in triggering the
bulk of the luminous infrared emission.
Many observations have shown that starbursts, induced by the
gravitational interaction of merging galaxies, could account for the
infrared emission (see Sanders \& Mirabel 1996 for a review). 
On the other hand, 
some observations have shown that active galactic nuclei (AGNs) might be 
the dominant energy source for some of the most powerful LIGs 
(e.g. Sanders {\it et al.} 1988; Genzel {\it et al.} 1998).

Numerical simulations of merging gas-rich spiral galaxies have shown how
the gas is redistributed during the merging process
(e.g., Barnes \& Hernquist 1996; Mihos \& Hernquist 1996).
By assuming that the star formation rate (SFR) was proportional to some power
of density (Schmidt Law), Mihos \& Hernquist (1996) inferred that 
sharp changes of the SFR of the galaxies occur 
as merging progresses. 
This simulated SFR merely reflects 
the density evolution of the gas. 
Direct observational evidence is necessary in order to 
establish how and when starbursts are initiated during the merging process. 

We have undertaken a study consisting of 
millimeter, optical, radio, and 
mid-infrared observations of an entire 
merging sequence of LIGs in an attempt to trace the evolution
of gas distribution and starbursts. A sample of gas-rich
LIGs was chosen to represent different phases of the interacting/merging
process, with emphasis on the the early and
intermediate merging phases so that the physical state of the ISM 
leading to starbursts
can be identified.  
In this {\it Letter}, we present results of mid-infrared images of these LIGs 
observed with the ISOCAM on the Infrared Space Observatory ({\it ISO\/}).
The purpose of the mid-infrared imaging is to locate the  
source and distribution of star formation with fairly good resolution 
which is unavailable at far-infrared.
    
\section{OBSERVATIONS AND DATA REDUCTION}

We carried out the mid-infrared observations of a sample of  
galaxies using the ISOCAM imaging facility (Cesarsky {\it et al.} 1996)
onboard the {\it ISO\/} satellite (Kessler {\it et al.} 1996).
All these galaxies have far infrared luminosities 
$L_{FIR} \ge 2 \times 10^{11} L_{\sun}$ detected by {\it IRAS\/} 
and  were selected to represent different stages of a merging process.
These galaxies were observed with the LW3 (12 - 18 $\mu$m)
broad band filter, and when possible 
some of the galaxies were also observed with the
LW2 filter (5 - 8.5 $\mu$m) in order to have a control on the
contamination of polycyclic aromatic hydrocarbon (PAH) emission on the
continuum flux. 
Images of these galaxies were obtained in a $2 \times 2$ raster mode with 
the 3$\arcsec$ pixel field of view (PFOV) lens and a step size of  
48$\arcsec$; the resulting field of view is $\sim 144\arcsec$.
We conducted thirteen exposures of 5.04-second integration 
per raster position for LW3 and LW2 observations.
Ten additional exposures were made before these observations
for stabilization.
 
The data reductions were performed using the CAM Interactive Analysis 
(CIA)
and softwares developed at IPAC. The data were deglitched 
using the multi-resolution median transform (MMT) filtering model 
to remove spurious spikes caused by cosmic rays 
(Siebenmorgen {\it et al.} 1996); 
the data were then checked and deglitched 
manually as necessary. 
Detector transients were fitted and removed using the 
IPAC simplified analytic model. 
To do dark and flat field corrections, we 
used the flat fields estimated from the median of all frames
and adopted the dark current images 
from the calibration libraries.
The resulting data were then mosaiced to produce the final sky maps.

\section{RESULTS AND DISCUSSIONS}

The ISOCAM LW3 and LW2 images of these LIGs are displayed in 
Figures 1 \& 2. 
The mid-infrared emission 
of these LIGs is extended with respect to the 
resolution of the beam size even for the most advanced merger
UGC 2369-S
(the southern galaxy of UGC 2369). 
The emission regions extend more than a few 
kiloparsec in these merger systems, indicating that the 
heating sources of the mid-infrared emission
also extend over a similar size scale. 

The 7 $\micron$ emission (Figure 2) is dominated by PAH emission.
For comparison we summarize the measured LW2 and
LW3 fluxes of our observations 
in Table 1. 
We note that the flux density ratios $S_{\nu}(15 \micron)/S_{\nu}(7 \micron)$
are $\sim 2$ 
for the early merger ARP 302
and the northern galaxies of UGC 2369 (UGC 2369-N), 
while the flux ratios are much higher ($\sim$ 3--5) 
for the more advanced mergers.
Since the 7 $\micron$ waveband is dominated by the PAH features 
(Acosta-Pulido {\it et al.} 1996), our results indicate that the 
relative strength of the PAH emission might decrease in
advanced merging stages. 

Our results exclude active galactic nuclei 
and favor a star-formation origin for the
mid-infrared emission because of their extended emitting sizes. 
The spatial distribution of infrared emission from these LIGs 
is different from
that of ultra-luminous infrared galaxies (ULIGs), which often
show point-source emission at mid-infrared wavelengths
(e.g. Sanders {\it et al.} 1988; Genzel {et al.} 1998)
and are thought to host an AGN as the dominant source for the 
infrared luminosity.
Nontheless, 
it cannot be ruled out that an AGN might still
contribute to the total luminosity for some of these LIGs.
In fact, some LIGs in our sample have been identified as Seyfert 2 
galaxies (e.g. NGC 5256 and NGC 7592) and some portion of 
their infrared luminosity
emission might come from a central AGN.  

The extension of the 15 $\micron$ emission is strongly
correlated with the apparent separation between the merging galaxies.
To investigate the possible concentration of the mid-infrared emission
in the merging process, we 
use the ratio of the peak to total flux of 
the 15 $\micron$ (LW3) emission as an indicator. 
In Figure 3a we plot the the peak-to-total-flux ratio as a function of  
projected separation between the two 
nuclei of these interacting galaxies.
The peak-to-total-flux ratios show a strong anti-correlation with
the separation; the linear correlation coefficient
is $-0.80$, corresponding to more than 95\% significance.    
For point sources, the ratios only reflect the distribution 
of the point 
spread function of the imaging facility and should be 
similar for any point source.  
The theoretical peak to total flux ratio of the point spread function 
is about 0.33 with the 3$\arcsec$ PFOV for the LW3 filter; 
it is obvious that our sample has a more extended distribution than a point 
source even for the most advanced merger UGC 2369-S. Furthermore,
we found that the ratios can be well fitted with a power law:
$$P/T=AD^{-p} ,$$
where $P/T$ is the peak-to-total-flux ratios of these LIGs, $D$ is the 
separation distances, and $p$ is the power-law index. 
The best-fit index is found to be $\sim$ 0.56. 
A simple interpretation for the correlation is that 
the mid-infrared emitting components 
(dust and molecular gas) of 
individual galaxies coalesce together as the merging progresses,
and the coalesced gas would have 
a higher gas density in a smaller volume, 
which would produce a higher star formation rate 
and a higher peak-to-total flux ratio if Schmidt Law is applicable.
If the behavior of the infrared emission distribution can be extrapolated 
to even more advanced mergers, these advanced mergers would have 
emission with a compact distribution similar to that of ULIGs.

Figure 3b and 3c plot the flux density
ratios $S_{\nu}(12 \micron)/S_{\nu}(25 \micron)$ and
$S_{\nu}(25 \micron)/S_{\nu}(60 \micron)$ of {\it IRAS} observations  
as a function of separation distances for these interacting galaxies.
The $S_{\nu}(12 \micron)/S_{\nu}(25 \micron)$ ratios decrease as the  
separation become smaller. Since the {\it IRAS} 12 $\micron$
emission is dominated by PAH features, this result indicates that 
the relative PAH strength decreases as the galactic interaction progresses.
This is consistent with our measurements of the ISOCAM 7 $\micron$ 
to 15 $\micron$ flux density ratios.
We note that Genzel {\it et al.} (1998) have also found a  
a similar trend for the 7.7 $\micron$ PAH emission 
among starburst galaxies, ULIGs, and AGNs.  
On the other hand, 
the $S_{\nu}(25 \micron)/S_{\nu}(60 \micron)$ 
ratios increase as the separation distance become smaller.
These results indicate that the inter-stellar radiation field (ISRF) that 
would heat the dust and destroy the PAH molecules in the galaxies 
become stronger as the merging progresses.
This is consistent with the star formation rates increasing as the galaxies
merge. The representative values of $S_{\nu}(25 \micron)/S_{\nu}(60\micron)$
for a galaxy hosting an AGN is usually $>$ 0.18 (Helou 1986); however,
the highest $S_{\nu}(25 \micron)/S_{\nu}(60 \micron)$
in our sample is only $\sim$ 0.18,
indicating that AGN emission is not important
for these LIGs.

\acknowledgments
The ISOCAM data presented in this paper was analysed using "CIA",
a joint development by the ESA Astrophysics Division and the ISOCAM
Consortium led by the ISOCAM PI, C. Cesarsky, Direction des Sciences de la
Matiere, C.E.A., France. 
We thank K. Ganga and the ISO staff at IPAC for their help.
KYL and RG were supported in part by an NASA/ISO grant 961504,
administered through the Jet Propulsion Laboratory of the California
Institute of Technology. CYH and KYL  
acknowledge support from the Academia Sinica and the National Science Council
of the Republic of China in Taiwan. 
YG, RG and KYL also acknowledge support from the Laboratory of
Astronomical Imaging which is funded by NSF grant AST 96-13999 and by the
University of Illinois.

\newpage
\begin{deluxetable}{lccccc}
\tablewidth{0in}
\tablecaption{ISO Observations of Luminous Infrared Galaxies.}
\tablehead{
\multicolumn{1}{l}{Source} &
\multicolumn{1}{c}{Separation (kpc)} &
\multicolumn{2}{c}{LW3 Filter (15 $\mu$m)} &
\multicolumn{2}{c}{LW2 Filter (7 $\mu$m)}\\
\colhead{} &
\colhead{} &
\colhead{$S_{T}$ (Jy)\tablenotemark{a}} &
\colhead{$S_{P}$ (Jy)\tablenotemark{a}} &
\colhead{$S_{T}$ (Jy)\tablenotemark{a}} &
\colhead{$S_{P}$ (Jy)\tablenotemark{a}}}
\startdata
ARP 302 & 25.8 & 0.51 $\pm$ 0.01 & & 0.27 $\pm$ 0.01 &  \nl
(VV 340A) & & 0.40 $\pm$ 0.01 & 0.024 $\pm$ 0.001 & 0.22 $\pm$ 0.01 &
0.015 $\pm$ 0.001 \nl
(VV 340B) & & 0.11 $\pm$ 0.01 & 0.0049 $\pm$ 0.0007 & 0.053 $\pm$ 0.003 &
0.0034 $\pm$ 0.0016 \nl
NGC 6670 & 14.6 & 0.56 $\pm$ 0.01 & & & \nl
(NGC 6670-E)\tablenotemark{b} & & 0.29 $\pm$ 0.01 & 0.023 $\pm$ 0.001 & & \nl
(NGC 6670-W)\tablenotemark{b} & & 0.27 $\pm$ 0.01 & 0.018 $\pm$ 0.001 & & \nl
UGC 2369 & & 0.70 $\pm$ 0.01 & & 0.21 $\pm$ 0.01 & \nl
(UGC 2369-S)\tablenotemark{c} & 2.2 & 0.65 $\pm$ 0.01 & 0.11 $\pm$ 0.01 &
0.18 $\pm$ 0.01 & 0.028 $\pm$ 0.001 \nl
(UGC 2369-N) & & 0.053 $\pm$ 0.003 & 0.0023 $\pm$ 0.0005 &
0.029 $\pm$ 0.002 & 0.0022 $\pm$ 0.0006 \nl
NGC 7592 & 7.1 & 0.66 $\pm$ 0.01 & 0.054 $\pm$ 0.001 &  &  \nl
NGC 5256 & 5.5 & 0.56 $\pm$ 0.01 & 0.054 $\pm$ 0.002 &  & \nl
MRK 848 & 4.8 & 0.52 $\pm$ 0.01 & 0.056 $\pm$ 0.001 & 0.10 $\pm$ 0.01 &
0.014 $\pm$ 0.004 \nl
NGC 6090 & 3.5 & 0.53 $\pm$ 0.01 & 0.058 $\pm$ 0.001 &  0.15 $\pm$ 0.014 &
0.020 $\pm$ 0.001 \nl
\enddata
\tablenotetext{a}{$S_{T}$: total flux density; $S_{P}$: peak flux density.}
\tablenotetext{b}{NGC 6670-E: the eastern galaxy of NGC 6670;
NGC 6670-W: the western galaxy of NGC 6670.}
\tablenotetext{c}{The southern galaxy of UGC 2369 shows double
nuclei at near-infrared and is assigned as an advanced merger.}
\tablenotetext{d}{The errors shown in the table are statistical errors;
the systematic errors might be larger than 10 \% (Ganga 1998).}
\end{deluxetable}

\newpage
\figcaption[figure1.ps]{ISOCAM LW3 
(15 $\micron$, ${\Delta}{\lambda} = 2.75 \micron$)
images of seven LIGs in a merging sequence. The sequence ordered from 
early to advanced mergers 
is Arp 302, NGC 6670, NGC 7592, NGC 5256, MRK 848, NGC 6090,
and UGC 2369-S. The last image shows the point spread function (PSF) of 
the ISOCAM LW3 filter for comparison. 
The size of each image is $75\arcsec \times 75\arcsec$.}   

\figcaption[figure2.ps]{ISOCAM LW2 
(7 $\micron$, ${\Delta}{\lambda} = 1.65 \micron$)
images of four LIGs in a merging sequence.
The sequence ordered from
early to advanced mergers
is Arp 302, MRK 848, NGC 6090, and UGC 2369. 
The size of each image is $75\arcsec \times 75\arcsec$.}

\figcaption[figure3.ps]{Infrared flux density ratios  
for LIGs in a merging sequence as a function of galaxy separation. 
(a): The peak-to-total 
flux density ratios derived from ISOCAM LW3 (15 $\micron$) images (square)
and 
a power-law fit for the ratios as a function of galaxy separation (solid line).
(b): The ratios of the {\it IRAS} 12 to 25 $\micron$ flux densities.
(c): The ratios of the {\it IRAS} 25 to 60 $\micron$ flux densities.    
} 

%\setcounter{figure}{0}
%\newpage
%\begin{figure}
%\plotone{fig1.ps}
%\caption{}
%\end{figure}
%\newpage
%\begin{figure}
%\plotone{fig2.ps}
%\caption{}
%\end{figure}
%\newpage
%\begin{figure}
%\plotone{fig3.ps}
%\caption{}
%\end{figure}
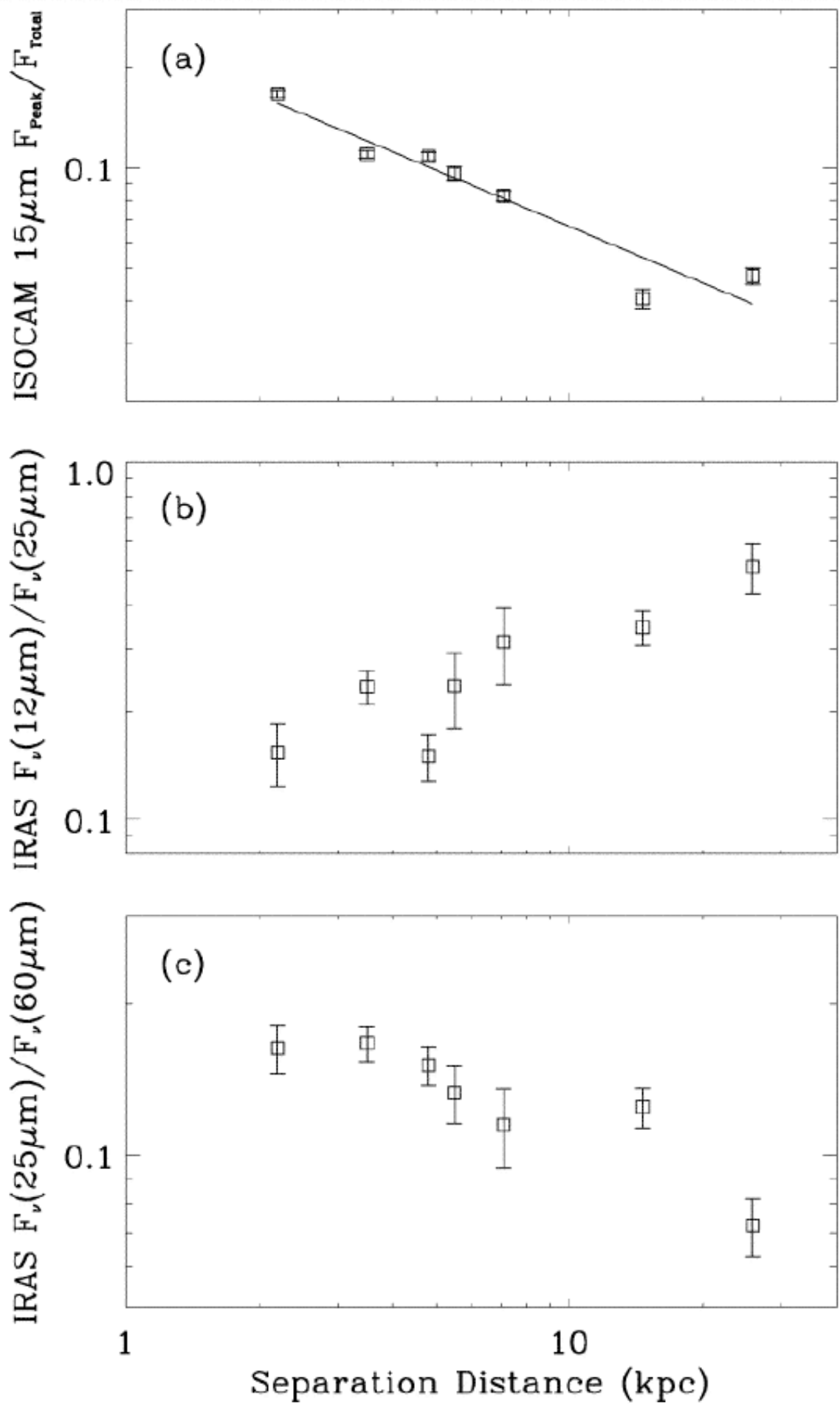
\end{document}